\documentclass[]{vgtc} 

\onlineid{1022}

\title{Toward Supporting Narrative-Driven Data Exploration: Barriers and Design Opportunities}

\author{%
  \authororcid{Oliver Huang}{0009-0007-1585-1229}\thanks{e-mail: oliver@cs.toronto.edu}\\ %
        \scriptsize University of Toronto %
  \and
  \authororcid{Carolina Nobre}{0000-0002-2892-0509}\thanks{e-mail: cnobre@cs.toronto.edu}\\ %
        \scriptsize University of Toronto %
}

\abstract{%
  Analysts increasingly explore data through evolving, narrative-driven inquiries, moving beyond static dashboards and predefined metrics as their questions deepen and shift. As these explorations progress, insights often become dispersed across views, making it challenging to maintain context or clarify how conclusions arise. Through a formative study with 48 participants, we identify key barriers that hinder narrative-driven exploration, including difficulty maintaining context across views, tracing reasoning paths, and externalizing evolving interpretations. Our findings surface design opportunities to support narrative-driven analysis better.
}


\teaser{
  \vspace{-20pt}
  \centering
  \includegraphics[width=\linewidth, alt={}]{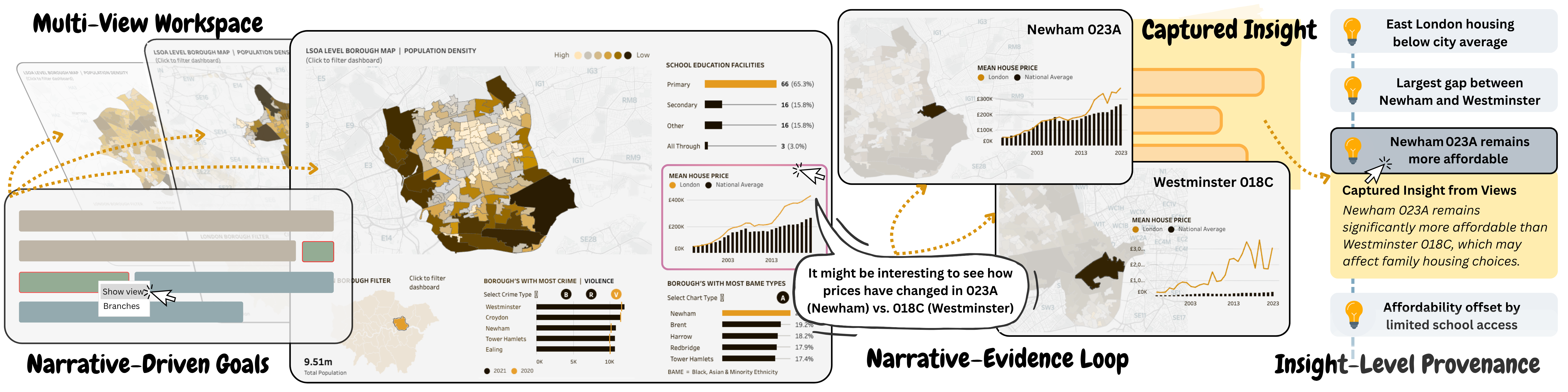}
  \vspace{-18pt}
  \caption{%
  Illustrated proposed workflow of narrative-driven exploration: (1) Analysts begin by defining narrative-driven goals, capturing evolving lines of inquiry. (2) They explore multi-view workspaces to examine data from multiple perspectives. (3) Through the narrative-evidence loop, analysts capture insights by either annotating emerging observations or generating linked views from narrative statements. (4) Each captured insight is a semantic reasoning unit that links narrative and evidence. (5) Each insight is logged in an insight-level provenance trail, preserving context and reasoning continuity.%
  }
  \label{fig:teaser}
}

\renewcommand{\manuscriptnotetxt}{}

\usepackage{tabu}                      
\usepackage{booktabs}                  
\usepackage{lipsum}                    
\usepackage{mwe}                       
\usepackage{float}
\usepackage{amssymb}  

\usepackage{mathptmx}                  

\begin{document}


\firstsection{Introduction}
\maketitle
\vspace{-3pt}
Narrative-driven exploration empowers analysts to pursue data inquiries that evolve, branch, and adapt as new patterns emerge. Analysts fluidly move between questions, connect observations, and refine interpretations, following the natural flow of discovery. This approach pushes beyond the boundaries of rigid, predefined paths, which are designed for monitoring known metrics or answering narrowly scoped questions.

However, current analytic tools often fall short of supporting this dynamic process. Dashboards and programming notebooks typically enforce linear interactions, where insights are captured as isolated filters, charts, or code cells~\cite{setlur_framework}. Even mixed-initiative systems with conversational interfaces are primarily reactive~\cite{gap_batch}. They focus on single-turn summaries rather than preserving the reasoning flow that links evolving questions together. As a result, analysts who pursue narrative-driven approaches become fragmented across views, logs, and external notes, making it challenging to maintain context, reflect on discoveries, or communicate findings as coherent narratives.

To better understand how analysts progress from surface-level monitoring to narrative-driven exploration, we conducted a formative study with 48 participants. Most analysis begins with low-level tasks such as checking metrics, retrieving known values, and making simple comparisons. As their questions evolve, analysts naturally try to weave observations into a story by connecting insights, explaining causes, and drawing implications. These activities reflect higher cognitive engagement and mark the transition toward narrative-driven exploration. However, current analytic tools disrupt this progression, leaving reasoning fragmented, context lost, and analysts reliant on external workarounds. Our study reveals key barriers behind this gap and highlights design opportunities for scaffolding evolving, narrative-driven analysis.

\vspace{-5pt}
\section{Narrative Engagement Gaps and Barriers}

\vspace{-3pt}
To investigate challenges in exploratory analysis and the reflection process, we conducted a formative study with 48 expert dashboard users recruited via Prolific\footnote{\url{https://www.prolific.com}}. Participants (26 male, 22 female; ages 21–55) reported using interactive analytic tools weekly in professional roles such as data analysts (16), managers (7), healthcare professionals (6), and sales consultants (3). Each completed a 30-minute Qualtrics survey comprising semi-structured open-ended questions on goals, challenges, reasoning strategies. Participants were compensated \$8 USD.
\vspace{-5pt}
\subsection{Evolving Questions and Exploration Goals}
\vspace{-3pt}
To explore how users engage with data exploration beyond surface-level monitoring, we asked participants to describe the questions they typically aim to answer, how those questions evolve during use, and what forms of reasoning they wish current analytic tools would better support.
We analyzed their responses thematically, extending prior models of exploratory engagement~\cite{setlur_framework}. Two researchers independently coded the responses, resolving discrepancies through discussion to ensure reliability.

From this analysis, we identified six levels of cognitive engagement that reflect increasing exploratory intent: from reviewing basic metrics to constructing coherent, narrative-style interpretations that span multiple layers of abstraction. Table~\ref{tab:levels} presents the number of participants who reported reaching each level in practice versus those who aspired to reach it when needed.

While most participants reported operating at the lower levels, such as monitoring metrics or making simple comparisons, their aspirations leaned heavily toward higher-level reasoning. Many wanted to connect insights across views, explain relationships, and structure their findings into stories that could be shared or revisited. As one participant explained:\textit{"Sometimes I start with a simple KPI check, but then I want to understand why something changed… That’s when I need to create new views, switching to spreadsheets or taking screenshots just to piece things together."}(P12). Another participant echoed this challenge from a frustration perspective: \textit{"As I explore, I keep jumping between charts and views, and it’s hard to keep my thoughts connected as a single story."}(P30). These perspectives reveal a gap between analysts’ evolving goals and the support provided by current tools. While analysts naturally progress from simple checks toward connecting insights into coherent stories, their reasoning often becomes fragmented and difficult to maintain.
\vspace{-10pt}

\renewcommand{\arraystretch}{1.4} 
\begin{table}[H]
  \caption{%
  	 Levels of cognitive engagement in data exploration, from surface monitoring (Level 0) to structuring findings into coherent stories (Level 5). Responses reveal a strong aspiration toward deeper reasoning and narrative-style exploration.%
  }
  \vspace{-0.8em}
  \label{tab:levels}
  \scriptsize
  \centering
  \begin{tabu}{%
    X[0.6,l]  
    X[9.4,l]  
    @{~~~}c  
    @{~~~}c  
  }
  \toprule
  \textbf{Lvl} & \textbf{Definition} & \textbf{Act.} & \textbf{Des.} \\
  \midrule
  0 & Reviewing surface-level summaries without interpretation & 7 & 0 \\
  1 & Retrieving known values or simple answers & 14 & 1 \\
  2 & Comparing data across categories or timeframes & 21 & 6 \\
  3 & Investigating relationships and forming tentative explanations & 3 & 23 \\
  4 & Connecting patterns across domains and exploring trade-offs & 1 & 7 \\
  5 & Constructing and externalizing a coherent narrative of insights & 1 & 11 \\
  \bottomrule
  \end{tabu}
\end{table}
\renewcommand{\arraystretch}{1} 

\vspace{-15pt}

\subsection{Barriers to Narrative-Driven Exploration}
\vspace{-3pt}
To uncover why a gap exists between participants’ actual and desired levels of engagement, we analyzed their open-ended responses in greater detail. By grouping participants by their actual engagement levels, we examined how their reported challenges aligned with their stage of exploration. This analysis revealed four barriers that explain why many analysts remain at lower levels and struggle to advance toward narrative-driven exploration.

\textbf{Comparing insights across multiple views is challenging.} Participants often mentioned that answers live in \textit{“separate charts”} or \textit{“different tabs.”} To see relationships, they must duplicate views or copy screenshots into slide decks. These manual workarounds impose a high cognitive load, interrupt the analytic flow, and hinder the recognition of patterns that span perspectives.

\textbf{Limited support for iterative hypothesis-building.}
Respondents also mentioned that they can spot anomalies, but have no place to store tentative explanations while they test them. Current tools capture clicks and filters but not the analyst’s evolving interpretation; hypotheses end up in notebooks, sticky notes, or are forgotten. Without a way to branch, revise, and revisit these ideas, observations stay disconnected and never crystallize into the explanatory narratives characteristic of higher-level engagement.

\textbf{Difficulty maintaining and externalizing reasoning paths.} Participants who adapted to a narrative-driven exploration style emphasized a strong need to retain context and keep track of their evolving reasoning.
Many reported \textit{“losing the thread”} after applying multiple filters or drill-downs and lacked in-context mechanisms to record partial insights or intermediate interpretations. Some tools capture only interaction-level provenance, such as clicks or filter histories, but this does not preserve the semantic reasoning or interpretive logic necessary for narrative-driven exploration.

\textbf{Confirmation bias reinforces incomplete narratives.} Participants who engaged in narrative-focused reasoning noted that following an emerging story can lead to confirmation bias if alternative perspectives are overlooked. Current tools rarely suggest alternative perspectives, and analysts can prematurely lock into a single narrative, risking overconfidence and missed insights.

\vspace{-5pt}
\section{Design Opportunities}

\vspace{-3pt}
To address the barriers identified in our study, we propose four design opportunities to support users’ unmet needs during narrative-driven exploration directly. 
\\\textbf{Enable Seamless Multi-View Synthesis:}  
Participants reported difficulty synthesizing insights when information was fragmented across different views. Analytic tools should support flexible, customizable multi-view layouts to reduce cognitive overhead and help analysts interpret cross-view relationships more easily. Seamless multi-view comparison will help analysts smoothly transition from isolated observations to integrated, narrative-level insights.
\\\textbf{Scaffold Iterative Hypothesis-Building:}  
Narrative-driven exploration requires structured ways to capture and refine evolving interpretations, which leaves tentative hypotheses external to the exploration environment. Therefore, tools should provide mechanisms to record, revise explicitly, and branch hypotheses during analysis. This way,  users can iteratively test explanations and track evolving ideas to align exploration with narrative practices.
\\\textbf{Preserve Reasoning Paths with Insight-Level Provenance:}  
Narrative-driven exploration depends on maintaining context, continuity, and a clear trail of analytic reasoning. To encourage deeper reflection and storytelling, analytic systems should maintain clear reasoning paths by incorporating features such as integrated annotations, insight checkpoints, and rationale-level provenance tracking. Making these reasoning steps explicit allows users to revisit and refine their thought processes more effectively.
\\\textbf{Promote Balanced Narrative Exploration to Reduce Bias:}  
Participants engaging in narrative-driven exploration noted that this practice carries risks of confirmation bias, because it encourages focusing primarily on evidence that fits existing narratives. Tools should actively help analysts identify and consider alternative perspectives or contradictory evidence during exploration. By prompting reflective checks or surfacing diverse views, analytic systems can ensure that emerging narratives remain balanced, rigorous, and comprehensive.

\vspace{-7pt}
\section{Future Work: Toward a New Workflow}
\vspace{-3pt}

While prior systems capture only interaction-level provenance and treat history as a retrospective artifact, they provide little support for the transition from non-linear exploration to linear narrative construction~\cite{madanagopal2019analytic}. We propose \textbf{narrative scaffolding} as a new workflow that treats interpretation as a dynamic, evolving process rather than a post-hoc activity. By embedding reasoning capture into the exploration environment, narrative scaffolding enables analysts to retain context, record evolving hypotheses, and externalize reasoning paths, allowing narratives to emerge naturally instead of being laboriously assembled after the fact. This proposed paradigm directly addresses our four design goals: enabling multi-view synthesis, scaffolding iterative hypothesis-building, preserving semantic provenance, and promoting balanced, bias-aware exploration.

Our next steps focus on building a prototype environment that supports branching exploration paths, in-context annotations, and reasoning continuity. We plan to evaluate how narrative scaffolding helps analysts reflect, connect insights, and communicate shareable narratives without fragmenting their workflow. To support community engagement and future contributions, the source code is available at
\href{https://github.com/hivelabuoft/Visplora.git}{github.com/hivelabuoft/Visplora}.

\vspace{-10pt}

\bibliographystyle{abbrv-doi-hyperref}


\begin{thebibliography}{1}

\bibitem{gap_batch}
A.~Batch and N.~Elmqvist.
\newblock The interactive visualization gap in initial exploratory data analysis.
\newblock {\em IEEE Transactions on Visualization and Computer Graphics}, 24(1):278--287, Jan 2018.

\bibitem{madanagopal2019analytic}
K.~Madanagopal, E.~D. Ragan, and P.~Benjamin.
\newblock Analytic provenance in practice: The role of provenance in real-world visualization and data analysis environments.
\newblock {\em IEEE Computer Graphics and Applications}, 39(6):30--45, 2019.

\bibitem{setlur_framework}
V.~Setlur, M.~Correll, A.~Satyanarayan, and M.~Tory.
\newblock Heuristics for supporting cooperative dashboard design.
\newblock {\em IEEE Transactions on Visualization and Computer Graphics}, 30(1), Jan. 2024.

\end{thebibliography}

\end{document}